# A New Information Hiding Technique Matching Secret Message And Cover Image Binary Value.

G.Umamaheswari[1]

[1]Research Scholar, Manonmaniam Sundaranar Univ.,
Dept. of Computer Applications,
SSS Jain College For Women, Chennai, India.
gumaganesh2011@gmail.com.

Dr.C.P.Sumathi[2]

[2]Dept. of Computer Science,
SDNB Vaishnav College for Women, Chennai, India.
drcpsumathi@gmail.com.

*Abstract*— Steganography involves hiding a secret message or image inside another cover image. Changes are made in the cover image without affecting visual quality of the image. In contrast to cryptography, Steganography provides complete secrecy of the communication. Security of very sensitive data can be enhanced by combining cryptography and steganography. A new technique that uses the concept of Steganography to obtain the position values from an image is suggested. This paper proposes a new method where no change is made to the cover image, only the pixel position LSB (Least Significant Bit) values that match with the secret message bit values are noted in a separate position file. At the sending end the position file along with the cover image is sent. At the receiving end the position file is opened only with a secret key. The bit positions are taken from the position file and the LSB values from the positions are combined to get ASCII values and then form characters of the secret message.

*Keywords- Cover image, Least Significant Bit(LSB), Mean Square Error(MSE), Peak Signal to Noise Ratio(PSNR), Pixel Value Differencing(P VD).*

## I. INTRODUCTION

With the rise in sharing of information via the internet, it is very difficult to ensure that confidential information reaches safely from source to destination without anybody intercepting in the middle. Cryptography deals with hiding the content of the message, whereas steganography deals with hiding the very existence of the message itself. Steganography is a technique used for covert communication. Secret Information is hidden in a digital image. Steganography usually takes advantage of the limitations of HVS (Human Visual System). The main criteria behind steganography is imperceptibility and embedding capacity. A proper Steganographic algorithm should be chosen so that we achieve the best embedding capacity without compromising on the quality of the image. The objective of steganography is communication in a way that the true secret message is not visible to any outsider.

Image Steganography can be classified into two broad types according to the embedding domains. Embedding in the spatial domain such as LSB based approaches [1][2][3] , PVD Based approaches[4][5][6] and embedding in the transform domain such as [7 ]and [8].

Spatial Domain Embedding involves directly embedding the secret data in the pixel positions of the image. Least Significant Bit(LSB) is the most popular embedding scheme where secret data is embedded in the spatial domain of an image. Usually any changes made to the least significant bits of an image will not affect the quality of the image and will not be easily detectable.

LSB steganography [9] is described as follows:

If LSB of the pixel value $I(i,j)$ is equal to the message bit m to be embedded, $I(i,j)$ remain unchanged.

If not, the LSB of $I(i,j)$ is set to m.

The message embedding procedure can be described using Eqs. (1) (2) and (3) as follows,

$I_s(I,j)= I(i,j) – 1$   $LSB(I(i,j)=1)$ and m=0         (1)
$I_s(I,j)= I(i,j)$        $LSB(I(i,j)=)m$                (2)
$I_s(I,j)= I(i,j) + 1$   $LSB(I(I,j))\neq 0)$ and m=1      (3)

Transform Domain techniques are more flexible and involve less computations. Transform Domain Methods hide messages in significant areas of the cover image which makes them more robust against attacks [16].

## II. ANALYSIS & LIMITATIONS OF RELEVANT APPROACHES

### A. Strategies in Spatial Domain

In Weiqi Luo et. al's article Edge Adaptive Steganography-LSB Matching Revisited (For Gray Scale Images), the authors have expanded the LSB Matching by proposing an edge adaptive scheme. The embedding regions are selected according to the size of the secret message and the difference between the two consecutive pixels in the cover image. Sharper edge regions are used for lower embedding rates, more edge regions are released adaptively for data hiding [10].

Ms.G.S.Sravanthi et.al. analyze the performance of edge adaptive steganography for colored images. This paper also encrypts the message using efficient cryptographic algorithm which further increases the security. The author has suggested a Plane Bit Substitution Method(PBSM) in which message bits are embedded into the pixel values of the image. A Steganography transformation Machine (STM) for solving binary operation for manipulation of original image to least significant bit(LSB) operator based matching[11] is suggested.

Tanmay Bhattacharya et.al, in their paper Novel Session Based Text Encryption & Hiding Technique suggest a novel approach towards document hiding technique within a color bitmap image using cross fold transposition and genetic algorithm. The secret text is converted into its equivalent binary form upon which cross fold transposition is applied. This binary form is perturbed by genetically generated session-key and then is embedded within the host image[12].

Bawankar Chetan D et. al in Pattern Matching with External Hardware for Steganography Algorithm propose a symmetric model approach. It uses the text message itself for encrypting data. A Secure authentication technique for binary images using pattern matching is also proposed by authors. A hardware kit with microcontroller is used at both ends to steganize and de-steganize the media [13].

In Pixel Intensity Based Steganography with improved Randomness Venkata Abhiram et.al has suggested random pixel manipulation with adaptive strategies. The least significant bits of any one of the channels RED, GREEN, BLUE is used as a pointer to decide the embedding capacity in the other two channels.

### B. Strategies in Transform Domain

In the article Application of CL multi-wavelet transform and DCT in Information Hiding Algorithm, Tao Zhang et.al suggest an algorithm that takes advantage of Discrete Cosine Transform & CL multi-wavelet transform. Experimental results indicate that the proposed scheme can increase invisibility and robustness. The CL multi-wavelet transform can process multi-transformation at the same time. CL-DCT has excellent sensitivity against cutting to image attacks.[4].

H.S.Manjunatha Reddy et.al suggest a high capacity and secure steganography using the Discrete Wavelet Transform. The wavelet coefficients of both cover and payload are fused into a single image using embedding strength parameters alpha and beta. The capacity and security is increased with acceptable PSNR in the proposed algorithm[5].

Rupa Maan et. al has come up with an idea of implementing a Biometric steganography that uses skin region of images in DWT domain for embedding secret data. By embedding only in certain regions (here skin region) and not in the whole image, and so security is enhanced. It is observed that the algorithm preserves the histogram of DWT coefficients after embedding also[14].

Amitava Nag et.al in A Novel Technique For Image Steganography Based on DWT and Huffman Encoding has suggested a new method where the cover image is subjected to Discrete Wavelet Transform . Huffman Encoding is performed on the secret message and are embedded in the high frequency coefficients that resulted from the Discrete Wavelet Transform[15].

### III. PROPOSED ALGORITHM

After analysis of all the above approaches it is found that even a very small secret message embedded into a cover will make some changes in the visual and statistical data of an image and will affect the image quality. Thus the image becomes vulnerable to attacks. To avoid this a new approach that doesn't make any changes in the cover image but notes down the pixel positions that match with the secret data is proposed. The positions that match with the secret data are stored in a separate file and the file is opened only with a secret key that is known only to the sender and receiver.

### A. Pixel Matching Module

In this module the cover image is read and separated into RGB components. The RGB ASCII values are converted into binary. The secret data is read from a text file and is separated into individual characters. Each character is converted into its corresponding ASCII value and then into binary. The binary values are stored as a bit stream.

Each pixel position is analyzed in the cover image. If the $8^{th}$ bit of the pixel matches with the secret bit stream the pixel position of the cover image is noted in a separate file. The same is repeated until all the values in the bit stream are matched with a position in the cover image. In our proposed algorithm only the green component is chosen for pixel matching. If the secret contains many characters then we can also include the red and the blue component.

### B. Extraction Module

The position values are stored in a position file which is opened only if the secret key matches. The stego image is separated into its RGB components. The pixel values are converted into binary. The positions are taken from the text file. The 8$^{th}$ bit values from the matching positions are taken and stored in an array. A set of 7 bits are grouped to form an ASCII value and then converted into its corresponding character. The characters are combined and the secret message is obtained. If secret key does not match the text file is not opened and a wrong secret key message is displayed. Figure 1 and Figure 2 depicts the steps in embedding and extraction.

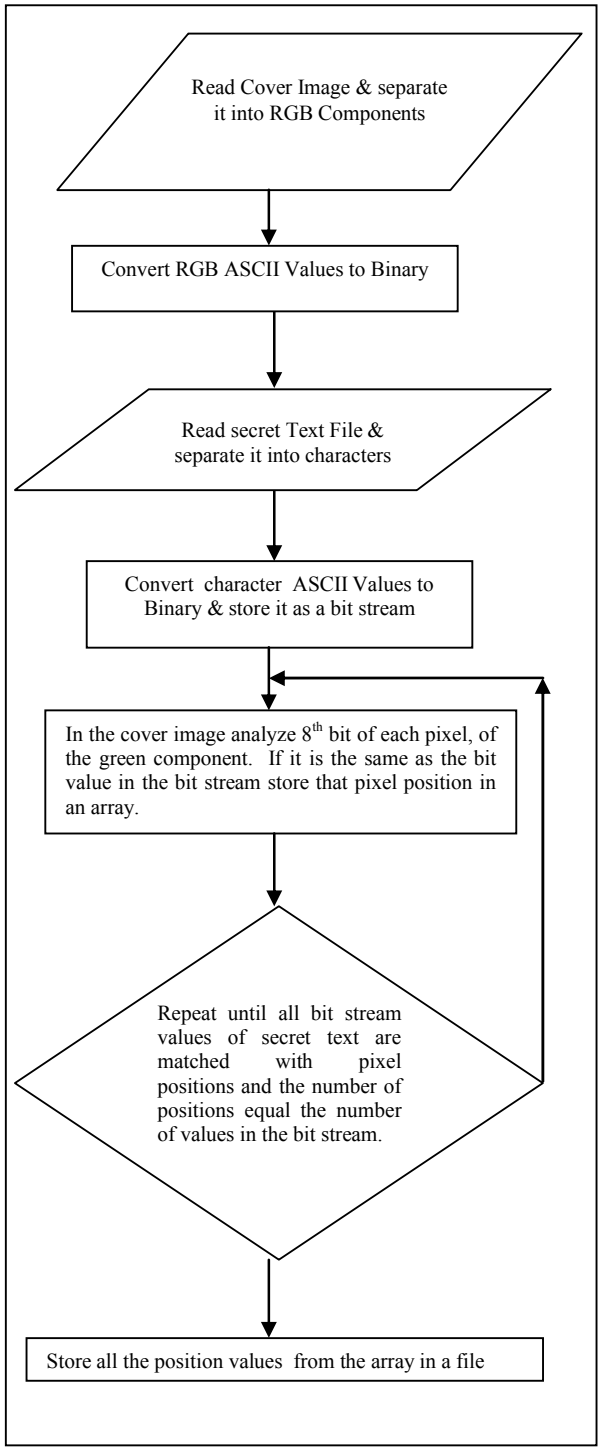

Figure 1 : Flowchart depicting the Embedding process

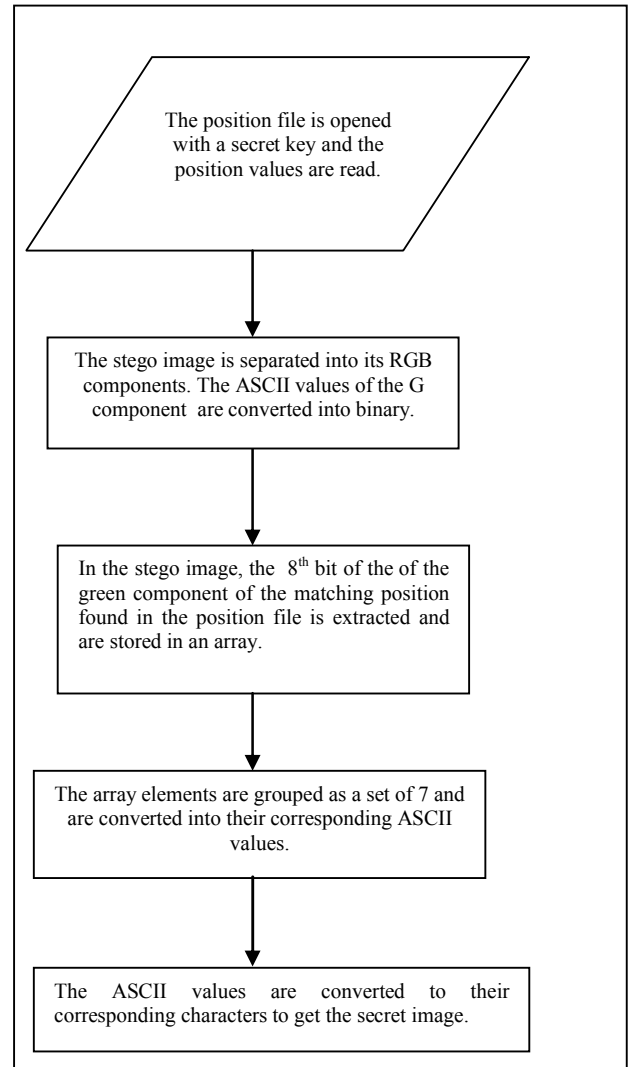

Figure 2 : Flowchart depicting the Extraction process

## IV. IMPLEMENTATION

For implementing the above algorithm MATLAB Version 7 .11.0 (R2010b) was used. Figure 3 shows the three images which includes Mozilla Firefox image of size 64 x 64, Lena image of size 256 x 256 and Babbon image of size 512 x 512.

The text "Hello World" is the secret message to be conveyed to the recipient. Instead of embedding the secret text in the image, we are matching the binary equivalent of the secret message characters with the binary values of an image. Only the LSB match is checked for noting the position values.

The secret message to be hidden is "HELLO WORLD". Therefore the 10 secret characters are equivalent to 70 binary digits.(Each characters ASCII value corresponds to a 7 bit binary value). Table 1 gives the ASCII and binary values of the secret message. Hence 70 bit positions are to be identified in the cover image that matches with the secret message binary value and the bit positions are to be noted in a separate position file.

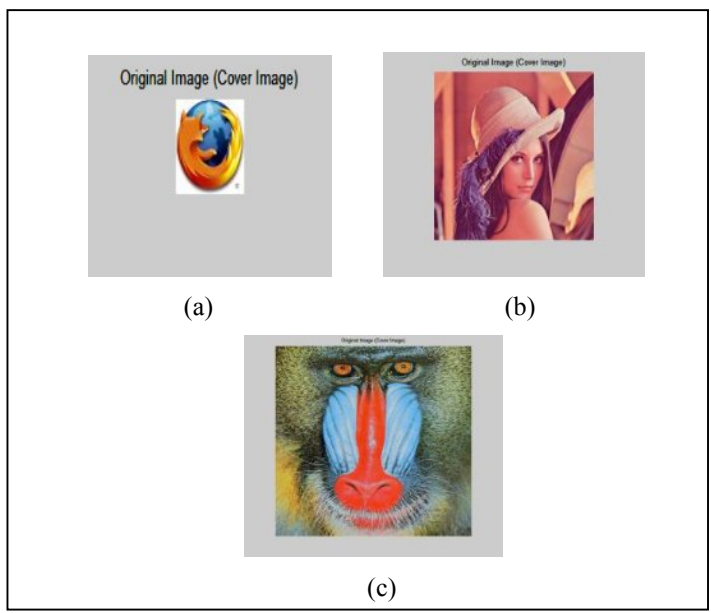

Figure 3: Sample Images (a) Firefox (b) Lena (c) Baboon

TABLE I : Secret Message Binary Values

| Character | ASCII Value | Binary Value |
|---|---|---|
| H | 72 | 1001000 |
| E | 101 | 1100101 |
| L | 108 | 1101100 |
| L | 108 | 1101100 |
| O | 111 | 1101111 |
| W | 87 | 1010111 |
| O | 111 | 1101111 |
| R | 114 | 1110010 |
| L | 108 | 1101100 |
| D | 100 | 1100100 |

The position file opens only with a secret key that is exchanged between the sender and the receiver prior to this communication. The secret image file and the position file can be sent separately so that anybody who comes across these files will not be able to infer anything from these individually.

If the position file is opened accidentally it will not reveal readable characters. The proper position integer values are got only through the extraction module in the matlab environment.

Also these files have utility only if used together. Tables 2, 3 & 4 gives the position values identified for the secret text.

TABLE II :- Positions Identified in Firefox Image

| | | | | | | | | | | | |
|---|---|---|---|---|---|---|---|---|---|---|---|
| 3 | 5 | 6 | 7 | 21 | 22 | 24 | 29 | 34 | 41 | 42 | 44 | 47 |
| 48 | 50 | 55 | 61 | 62 | 65 | 67 | 69 | 70 | 71 | 80 | 81 | 82 |
| 84 | 85 | 87 | 88 | 91 | 92 | 93 | 96 | 97 | 98 | 100 | 102 | 105 |
| 113 | 115 | 118 | 119 | 120 | 125 | 131 | 132 | 133 | 134 | 136 |
| 137 | 138 | 140 | 141 | 143 | 144 | 146 | 147 | 148 | 151 | 152 |
| 153 | 155 | 156 | 157 | 164 | 165 | 167 | 168 | 169 |

TABLE III :- Positions Identified in Lena Image

| | | | | | | | | | | | |
|---|---|---|---|---|---|---|---|---|---|---|---|
| 1 | 3 | 11 | 12 | 14 | 19 | 20 | 22 | 23 | 25 | 26 | 27 |
| 28 | 30 | 31 | 32 | 34 | 35 | 36 | 37 | 39 | 41 | 42 | 43 |
| 44 | 45 | 47 | 48 | 49 | 50 | 52 | 54 | 55 | 58 | 59 | 60 | 61 |
| 62 | 63 | 64 | 66 | 67 | 71 | 72 | 73 | 77 | 82 | 84 | 85 |
| 87 | 89 | 92 | 94 | 97 | 105 | 107 | 108 | 109 | 111 | 112 | 113 |
| 115 | 117 | 118 | 119 | 127 | 129 | 134 | 136 | 137 |

TABLE IV :- Positions Identified in Baboon Image

| | | | | | | | | | | | |
|---|---|---|---|---|---|---|---|---|---|---|---|
| 1 | 3 | 4 | 5 | 6 | 8 | 9 | 10 | 15 | 16 | 18 | 21 |
| 22 | 25 | 26 | 28 | 36 | 37 | 38 | 39 | 43 | 45 | 46 | 47 |
| 48 | 49 | 52 | 53 | 54 | 55 | 59 | 60 | 62 | 64 | 65 | 66 |
| 67 | 70 | 72 | 73 | 74 | 76 | 77 | 81 | 85 | 88 | 89 | 90 |
| 91 | 92 | 93 | 97 | 101 | 102 | 106 | 107 | 108 | 109 | 111 | 112 |
| 113 | 115 | 116 | 117 | 120 | 123 | 124 | 126 | 127 | 128 |

Since no change is made in the cover image any intruder will not be able to extract anything from the cover image alone. Even a study of the histogram will not reveal any changes as there will not be any change in the histogram of the cover and secret images (Figure 4 & Figure 5).

It is also found that the MSE (Mean Square Error) value is zero and the PSNR (Peak Signal to Noise Ratio) value is Infinity. This is because no change is made to the original image, only the matching position values are noted.

This algorithm is a generalized one which can be used with any image for identification of the positions. The only criteria to be noted is the size of the image to be taken with respect to the size of the secret message to be embedded.

To add another layer of security the position file can be encrypted and sent. In our implementation we have not encrypted the position file as it will not convey any meaningful information if opened directly. The positions can be extracted only through MATLAB environment. Also the image file and the position file can be sent separately. An image which is commonly available in the internet can also be chosen for this purpose.

Although the presented paper has tested with 10 characters, in actual case more number of characters can be matched with pixel values.

## V. PERFORMANCE ANALYSIS:-

The ususal measures for analyzing of the quality of the stego image are MSE (Mean Square Error), PSNR (Peak Signal to Noise Ratio) , Embedding Capacity and Robustness (Tolerance towards attacks).

MSE (Mean Square Error) is the cumulative squared error between the original and the stego image. PSNR measures the Peak signal to Noise Ratio that gives the quality of the image after embedding secret data[16]. Equations 4 and 5 represent the mathematical formulae for MSE and PSNR.

$$MSE = \sum_{x=0}^{M-1}\sum_{y=0}^{N-1} \frac{(I(x,y)-I'(x,y))2}{M*N} \quad (4)$$

$$PSNR = 20 * \log 10 \frac{255}{sqrt(MSE)} \quad (5)$$

Where M and N denote the total number of pixels in the horizontal and vertical dimensions of the image and I represents the original image and I' represents the stego image. When the MSE is value is very low then it means there is very little difference between the original and the stego image.

The Peak signal-to-noise ratio is used to evaluate the image quality. If PSNR value is high it means the image quality is good even after embedding.

A proper balance should be maintained in selecting a cover image with respect to the size of the secret data. The size of the cover image should be such that the quality of the image is not compromised.

The sample secret message "Hello World" when converted into binary has 70 bits. Therefore 70 bit positions have to be identified. If we take a 64 x 64 size image. It can be divided into three 64 x 64 R, G, B channels. We have 4096 positions in each of the RED, GREEN and BLUE channels, out of which we will have at least 1000 positions that may find a match with the secret message. We are actually using only 10% of the actual capacity for this 64 x 64 sized image.

In Table 5 the Mean Square Error (MSE) value is zero as there is no difference between the original and the stego image.We get the PSNR value as infinity since 1/zero is infinity. The embedding capacity is calculated approximately. The sample calculation of the embedding capacity is as follows . For eg. in a 64 x 64 image the total number of bit positions are 4096 x 3 (For Each RGB channels). If we consider that one fourth pixel positions will match the secret then it is 4096 x 3 / 4 = 3,072 i.e 3000 bits approximately. * denotes approximate values.

The no of bits for which positions can be found can be calculated in the same way for other images. Since each alphabetic character is formed by combining 7 bits and then converting them to binary and finally to character, the total number of characters that can be embedded is 3000/7 i.e approximately 420 characters. This is a considerable size that can be exchanged between the sender and receiver.

TABLE V :Performance Analysis Table

| Image | MSE | PSNR | Embedding Capacity | No. of Characters |
|---|---|---|---|---|
| Firefox 64 x 64 | Zero | Infinity | 3000 bits * | 420 |
| Lena 256 x 256 | Zero | Infinity | 49000 bits * | 7000 |
| Baboon 512 x 512 | Zero | Infinity | 190000 bits* | 27000 |

The extraction module ensures that the secret message is properly extracted from the positions found in the position file for the specific image. In this case the extracted characters are "HelloWorld". The extracted characters (i.e) the secret data is stored in a file secretextract64anew.

The analysis of histograms in Figure 4 and Figure 5 of the images before and after noting down the positions will be identical because we have not made any changes in the actual image but we have noted only the positions where the LSB's match with the secret image binary value.

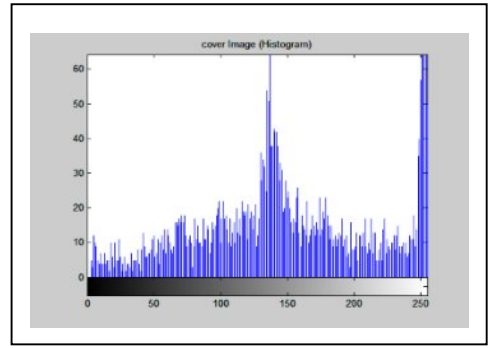

Figure 4: Histogram Before Position Identification

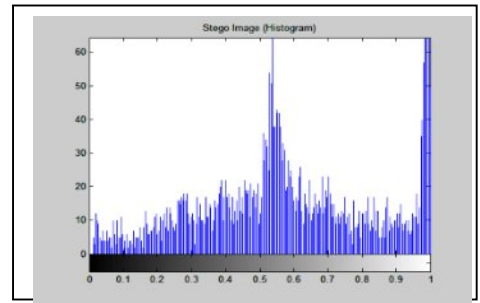

Figure 5: Histogram After Position Identification

## CONCLUSION

This paper has proposed a new perspective of exchanging secret data through steganography. This algorithm can be generalized by including the name of the image that is commonly available in the internet with the secret message and the positions for the name of the image included with the secret message. For example if the image is "Lena" and the secret message is "Hello World" the total letters for which positions have to be identified is 14 (HelloWorldLena). This can be extended to any image and any secret message.

### ACKNOWLEDGMENT

My heartfelt thanks to my guide who provided me great support and for the guidance given at appropriate times and for clearing my silly doubts.